\documentclass[twocolumn,preprintnumbers,amsmath,amssymb]{revtex4}
\usepackage{graphicx}

\begin{document}
\title{Baryogenesis within the two-Higgs-doublet model in the Electroweak scale }
\author{M. Ahmadvand}
\email{email: moslem.ph@gmail.com}
\affiliation{
Department of Physics, Shahid Beheshti University G.C., Evin, Tehran
19839, Iran
}%
\date{\today}

\begin{abstract}
The conventional baryogenesis mechanism is based on the one Higgs doublet within the standard model, at the electroweak scale $T\sim 100 GeV$. In this model the strong first order phase transition due to the spontaneous symmetry breaking imposes the folowing condition on the mass of the Higgs field: $m_H\lesssim 40 GeV$, which is contrary to the recently observed value $m_H\simeq 126 GeV$. In this paper we propose a baryogenesis mechanism within a two-Higgs-doublet model in which the phase transition occurs in one stage. This model is consistent with the observed mass of the Higgs. We obtain the true vacuum bubble wall velocity and thickness in this model. Then, we use nonlocal baryogenesis mechanism in which the interaction of fermions with the boundary of the expanding bubbles leads to CP violation and sphaleron mediated baryogenesis.
\end{abstract}
\maketitle

\section{Introduction}
The standard cosmology can approximately describe the history of the universe from very early times to the present time. One of the predictions of this theory is the relative abundances of the light nuclei produced in, primordial nucleosynthesis, through the input parameter $ \eta\equiv n_B/s $ where $ n_B $ is the difference between the number density of baryons and antibaryons in the universe and $ s $ is the entropy density in the universe\cite{kolb}. $ \eta $ corresponding to the observed abundances is given by \cite{astro}
\begin{eqnarray}\label{e00}
\eta\equiv\frac{n_B}{s}\simeq 10^{-10}.
\end{eqnarray}
After the discovery of maximal parity symmetry (P) violation in the weak interactions, many people thought CP operator is an exact mirror symmetry of the weak interaction. However, Cronin and Fitch found \cite{christenson} CP is violated in the neutral kaon system. Also the experiments \cite{gjesdal} showed that the long-lived neutral kaon decays more often into a positron than an electron by amount $ 3.3\times 10^{-10} $. This was the first discovery of a process that distinguished between the matter and antimatter and was linked to CP violation. Since then it has been thought that analogous processes could be responsible for preference matter over antimatter in the universe.\\ In  1967, Sakharov identified three conditions that are necessary for the generation of baryon asymmetry during the evolution of the universe from the symmetric inital state.\\ These conditions are:
 \begin{itemize}
 \item Violation of baryon number B: if baryon number is conserved in all interactions, no baryon asymmetry can be generatied from the symmetric initial state. 
 \item Violation of  C and CP: even if there exist B-nonconserving interactions, processes that produce excess baryons and antibaryons will have the equal rate. C and CP violation provide an imbalance.
 \item A departure from thermal equilibrium: in thermal equilibrium, the number density of baryons and antibaryons is given by $ [1+\exp((p^2+m^2)/T^2)]^{-1}  $. By CPT invariance, the masses of baryon and antibaryons are equal. Thereby, there is no net asymmetry produced. 
 \end{itemize}
 Although the standard model (SM) has all the necessary conditions, it cannot explain the baryon asymmetry since CP-violating effects arising through the Cabbibo-Kobayashi-Maskawa mechanism are too small \cite{cp}. Moreover, the ElectroWeak Phase Transition (EWPT) is strongly first order provided the mass of the Higgs boson is so lower than its observed value (see the reviews \cite{review, bern}).\\A number of extensions to the standard model have been proposed to overcome these shortcommings. The most popular extension of the standard model is supersymmetry (SUSY) \cite{susy}, in particular the minimal supersymmetric standard model (MSSM) \cite{mssm}. The importance of this model is due to solving the hierarchy problem, proposing candidates for dark matter and so on, in which a strongly first order phase transition can occur if the right-handed stop quark (super partner of the top quark) be light enough \cite{nmssm}. Also, the mechanisms are presented within the Two-Higgs-Doublet Model (THDM) in which the phase transition occurs in two stage \cite{land, schmidt}. In these models, the EWPT is weakly first order, and another first order phase transition occurs somewhat later. At this point, the Vacuum Expectation Value (VEV) in the additional Higgs direction changes to VEV in the direction the standard model Higgs. This tunneling process occurs with production of bubble walls and departure from thermal eqilibrium.\\ In this paper, we propose a baryogenesis mechanism within the THDM \cite{fromme}. This model includes two extra neutral and charged Higgs bosons beside the SM Higgs. We use the Higgs basis to explain EWPT and this phase transition occurs in one stage therefore we can explain EWPT by the SM Higgs dynamics. We find that phase transition occurs at temperature $ T_c \simeq 124 GeV $ and it is strong first order if the extra Higgs masses are about $ 221 GeV $. By this procedure, the strength of PT can vary from strong PT to very strong PT as a function of the mass of extra Higgs bosons. After phase transition bubbles form and expand which are required phenomena for baryogenesis at the electroweak scale. The boundaries or walls of the bubbles separate two phase states. Here, the interaction of fermion fields with the bubble walls causes CP violation. We compute the steady velocity of the wall in the thin wall aproximation and can apply nonlocal baryogenesis to calculate CP violation effects which enter the relevant equations.\\
 In section II we discuss baryon number violation in the electroweak (EW) theory and the rate of B-violating processes. Section III is devoted to CP violation in the EW theory and also in the THDM. Section IV contains the EWPT and its results in the THDM. In section V we obtain the $ \eta $ parameter using the nonlocal baryogenesis mechanism in the thin wall regime.

\section{Baryon number violation in the electroweak theory }
The first condition to explain the baryon asymmetry of the universe is violation of the baryon number B symmetry because, if no process which violates baryon number occurs, the total baryon number of the universe remains constant and no asymmetry can be generated from the symmetric initial conditions. There is baryon number violation in the SM which is completely negligible for particle reaction and collision energies in the labratories at the present time, but very significant for high energies in the early universe \cite{bern}. 
the SM lagrangian $ \cal L_{SM}$ with its strong interaction and electroweak parts is invariant under such global phase transformations of lepton fields $ \ell $ and quark fields $ q $ 
\begin{eqnarray}\label{e01}
\begin{array}{cc}q\rightarrow e^{i\theta /3}q  ,\hspace*{1cm}  &\ell\rightarrow \ell\\\\
\ell\rightarrow e^{i\theta} \ell ,\hspace*{1cm}   &q\rightarrow q \end{array}
\end{eqnarray}
at the classical level, the associated current $J^B_\mu , J^L_\mu$ are conserved
\begin{equation}\label{e02}
J^B_\mu=\frac{1}{2}\sum_q \overline{q}\gamma_\mu q ,\hspace{1cm} J^L_\mu=\frac{1}{2}\sum_\ell \overline{\ell}\gamma_\mu \ell . 
\end{equation}
However, at the level of quantum fluctuations these symmetries are explicitly broken because the price of requiring gauge invariance is the anomalous nonconservation of the axial current \cite{adler, jackiw} $J^{\mu5}=\overline{\psi}\gamma^\mu\gamma^5\psi  $ ,$ \psi =\{q,\ell \} $. 
This is relevant to baryon number since the EW fermions couple chirally to the gauge fields. We may write the baryon current as 
\begin{eqnarray} \label{e03}
J_\mu^B &=&1/4 [\overline{q}\gamma_\mu(1+\gamma^5)q+\overline{q}\gamma_\mu(1-\gamma^5)q]\nonumber\\\nonumber\\&=&1/2[\overline{q}_L\gamma_\mu q_L+\overline{q}_R\gamma_\mu q_R].
\end{eqnarray}
Only the axial current part is of importance. The Adler-Bell-Jackiw anomaly is given by
\begin{equation}\label{e04}
\partial_{\mu} J^{\mu 5}=\frac{-g^2}{32\pi^2}F_{\mu\nu}^i \widetilde{F}^{i\mu\nu} 
\end{equation}
where $\widetilde{F}^{i\mu\nu}=\varepsilon^{\mu\nu\alpha\beta}F_{\alpha\beta}^{i}$ is the (non)abelian field strength and g denotes the gauge coupling. The weak gauge bosons $ W^i_\mu (i=1,2,3)$ couple only to left-handed fermions $(\psi_L=q_L, \ell_L)$ , and the hypercharge boson couples to both $\psi_L $ and $ \psi_R $ with different strengths. Hence, we obtain
\begin{eqnarray}\label{e05}
\partial\mu J^\mu_B\ &=& \partial\mu J^\mu_L\nonumber\\ &=&n_f(\frac{g^2}{32\pi^2}G_{\mu\nu}^{i}\widetilde{G}^{i\mu\nu}-\frac{g'^2}{32\pi^2}B_{\mu\nu}\widetilde{B}^{\mu\nu})
\end{eqnarray} 
where $G^i_{\mu\nu}$ and $B_{\mu\nu}$ denote the $SU(2)_L $ and $ U(1)_Y$ field strength tensors, respectively. $n_f=3$ is the number of families, $g $ and $ g' $ are the  $SU(2)_L $ and $ U(1)_Y$ gauge coupling. Equation (\ref{e05}) may be written as
\begin{eqnarray}\label{e06}
\partial^\mu J^B_\mu &=& \partial^\mu J^L_\mu = n_f\partial_\mu K^\mu \\ K^\mu &=& \frac{g^2}{32\pi^2}\varepsilon^{\mu\nu\alpha\beta}(G^i_{\nu\alpha}W^i_\beta-\frac{1}{3}g\varepsilon_{ijk}W^i_\nu W^j_\alpha W^k_\beta)\nonumber\\&-&\frac{g'^2}{32\pi^2}\varepsilon^{\mu\nu\alpha\beta}B_{\nu\alpha}B_\beta 
\end{eqnarray}
$\partial_\mu K^\mu$ is the second Chern form and is gauge invariant. Chern forms can be locally written as exact forms. This expression constructed from the gauge fields whose exterior derivatives give the Chern form is called a Chern-Simons form, however none of these Chern-Simons forms are gauge invariant. $K^\mu$ is the second Chern-Simons form. If we integrate $\partial_\mu K^\mu$ over space-time, we will obtain the second Chern number 
\begin{equation}\label{e07}
\int d^4x\, \partial_\mu K^\mu =\frac{g^3}{96\pi^2}\int_s d^3x\, n_\mu \varepsilon^{\mu\nu\alpha\beta}\varepsilon_{ijk}W^i_\nu W^j_\alpha W^k_\beta . 
\end{equation}
Now, consider cylindrical space-time with ends at $ t_i $ and $ t_f $, and the volume of cylinder tend to infinity. Since $ \partial_\mu K^\mu $ is gauge invariant, we choose the temporal gauge condition, $ W^i_0=0 $. We separate spatial variables $ x $ and the time $ t $ since we are integrating differential forms over manifolds, the metric plays no role, so the time could be Euclidean or Mincowskian. There is no contribution from the boundary at spatial infinity because of our gauge choice.  
\begin{align}\label{e08}
\int d^3xdt\,\partial_\mu K^\mu =&n_{cs}(t_f)-n_{cs}(t_i)\\n_{cs}(t)=&\frac{g^3}{96\pi^2}\int d^3x\,\varepsilon_{abc}\varepsilon^{ijk} W^a_i W^b_j W^c_k
 \end{align}
where $ n_{cs}(t) $ is the Chern-Simons number.\\Equation (\ref{e08}) states that the integral of the second Chern form over space, and from time $ t_i $ to $ t_f $, is the change in Chern-Simons number. These numbers are the topological charges that classify the degenerate vacuum states. $ \Delta n_{cs} $ or the second Chern number is called instanton number. Instantons are topological solitons of pure Yang-Mills theory defined in four-dimensional Euclidean space-time and minimize the Euclidean action. They induce transitions that change $|i,t_i> $ to $|f,t_f> $ subject to $ \Delta B=\Delta L=n_f\Delta n_{cs} $. \\In the infinite dimensional gauge and Higgs field configuration space, vacua of the EW theory with different topological charges are separated by potential barriers. \\
Unstable solution of the field equations are usually minima of the energy within a subclass of fields with a certain symmetry, but saddle points of the energy in the space of the all field configurations. In the field theories, these saddle points are referred to as the sphalerons \cite{manton}. An (unstable) solution of the classical field equations of the SM gauge-Higgs sector is the sphaleron solution with Chern-Simons number $ 1/2 $.\\    
\begin{eqnarray}\label{e09}
E_{sph}(T)=\frac{4\pi}{g}\nu_T f(\frac{\lambda}{g}), 
\end{eqnarray}
where $ \nu_T $ is the vacuum expectation value of the SM Higgs doublet field at temperature T. The function $ f(\lambda /g) $ varies between $ 1.6<f<2.7 $ depending on the value of the SM Higgs mass \cite{klinkhamer}. We have $ \nu_{T=0} = 246 $ GeV and this gives $ 8\, TeV<E_{sph}(T=0)<14\, TeV $. In the zero temperature, the density of fermions and energy of colliding particles is low, thus baryon-number-violating processes occur through tunneling process between classical vacua. The tunneling amplitude between the degenerate vacua is proportional to $ \exp(-s_E)|_{inst} \, (M_{inst}\propto \exp(-s_E)  $ , where $ s_E $ is the pure gauge Euclidean action. The classical computaion of 't Hooft \cite{hooft} for the rate per unit volume of baryon-number-violating processes at zero temperature is approximately
\begin{equation}\label{e010}
\Gamma(T=0)\sim \exp(-2s_E)\sim 10^{-170}. 
\end{equation}
At high temperatures, when temperature becomes comparable with the barrier height, it becomes possible for termal transitions over the barrier to occur. The rate of these transitions will be proportional to $ \exp(-E_{sph}(T)/T)  $. In the phase where the EW gauge is broken, $ \nu_T \neq0  $ , the rate per unit volume of 
baryon-number-violating processes is given \cite{moore} 
\begin{equation}\label{e011}
\Gamma(T)=k_1(\frac{m_W}{\alpha_w T})^3 m_W^4 \exp(-E_{sph}(T)/T) 
\end{equation}
where $ m_W(T)={g\nu_T}/{2} $ is the temperature dependent mass of the W boson and $ k_1 $ is a dimensionless constant. The calculation of these processes in the unbroken phase is given by \cite{bodeker, moore2}
\begin{equation}\label{e012}
\Gamma(T)=k_2 \alpha_w^5T^4 
\end{equation}
with $ k_2=29\pm 6 $.\\Now, we shall discuss the second Sakharov condition.
\section{C and CP violation}
As mentioned earlier C and CP violation are necessary for baryogenesis scenario to proceed. In the SM, C symmetry is maximally violated since fermions in the EW theory are chirally coupled to the gauge fields and $ SU(2) $ gauge fields couple only with the left-handed fermions, but CP violation is a very small effect in the weak interactions.
\subsection{CP violation in the SM}
In the SM lagrangian, one cannot put explicit mass terms for fermions since the left- and right-handed fermion fields have different quantum numbers and so these terms violate gauge invariance. However, using the Yukawa couplings which are gauge invariant, one obtains fermion masses for the Dirac fields. For example, we write for quarks as 
\begin{align}\label{e013}
\cal L_Y=&-(\lambda_d^{ij} \overline{Q}^i_L\cdot  \Phi d_R^j+\lambda^{ij}_u \overline{Q}^i_L\cdot \widetilde{\Phi} u_R^j)+h.c., 
\end{align}
$ \lambda_u ,\lambda_d $ are general, not necessarily Hermitian, complex-valued matrices.
$ \Phi=\left( \begin{array}{c}\varphi ^+\\ \varphi^0 \end{array}\right) $ denotes the Higgs doublet and $ \widetilde{\Phi}\equiv i\tau_2 \Phi^{\dagger^T}  $ where $\tau_2=\left(\begin{array}{cc} 0&-i \\ i&0 \end{array}\right) $ and T denote the transpose operation.
We take the number of families of fermions to be $ n_f $. $ Q^i_L=\left(\begin{array}{c}u^i\\d^i\end{array}\right)_L $  where $ u^i=(u,c,t,...) $ and $  d^i=(d,s,b,...) $. If we substitude $ \Phi $ by its vacuum expectation value, we obtain the mass terms 
\begin{eqnarray}\label{014}
\cal L_M&=&-(M^{ij}_d \overline{d}^i_L d^j_L+M_u^{ij}\overline{u}^i_L u^j_R)+h.c.
\end{eqnarray}  
where $ M_d=\nu \lambda_d\, ,M_u=\nu \lambda_u $ are the mass matrices and $ \nu=\langle \varphi^0\rangle  $. These can be diagonalized by unitary transformations
\begin{eqnarray}\label{015}
\begin{array}{ccc}
u_L=U_L^u u'_L&,&u_R=U_R^u u'_R\\\\d_L=U_L^d d'_L&,&d_R=U_R^d d'_R
\end{array}
\end{eqnarray}
where $ u'_{L,R} \, , d'_{L,R} $ are the basis of quark mass eigenstates
\begin{eqnarray}
U_L^{u^\dagger} M_u U_R^u&=&\rm{diag}\,(m_u,m_c,m_t,...) \\\nonumber\\U_L^{d^\dagger} M_d U_R^d&=&\rm{diag}\,(m_d,m_s,m_b,...) 
\end{eqnarray}\label{016}
the matrices $ U^u $ and $ U^d $ cancel out of the pure kinetic terms as well as the electromagnetic and $ Z^0 $ boson currents. However, in the $ W $ boson current we have
\begin{eqnarray}\label{e017}
\cal L_W&=&\frac{g}{\sqrt{2}}\overline{u}'^i_L\gamma^\mu(U^{u^\dagger}_L U_L^d)^{ij}d'^j_L W_\mu +h.c. 
\end{eqnarray}
The matrix $ V= U_L^{u^\dagger}U_L^d$ is referred to as the Cabibbo-Kobayashi-Maskawa (CKM) mixing matrix \cite{CKM}. The matrix V is Hermitian and some of the phases in this matrix do not have physical meaning, in the sense that they can eliminated by rephasing the quark fields, only $ N=1/2(n_f-1)(n_f-2) $ parameters of  V remain as physical phases \cite{bra}. Thus, there is one physical phase in V when $ n_f=3 $. This phase is a source of CP violation, but is not strong enough to explain the baryon asymmetry of the universe. In the relevant formula in which the amount of CP violation is given, it torn out by a dimensionless constant which is about $ 10^{-20} $ and clearly this number is too small. Therefore, we attempt to add a further source in an extension of the SM.           
\subsection{the two-Higgs-doublet model (THDM)}
Most extensions of the SM expand the Higgs sector. To do this, there are many theoretical motivation.
One of the motivations is extra source of CP violation. In the THDM, the most general renormalizable scalar potential invariant under $ SU(2)\otimes U(1) $ is given by \cite{branco}
\begin{eqnarray}\label{e018}
V(\Phi_1,\Phi_2)&=&m_1|\Phi_1|^2+m_2|\Phi_2|^2+(m_3 \Phi_1^\dagger \Phi_2+h.c.)\nonumber\\&+&n_1|\Phi_1|^4+n_2|\Phi_2|^4+n_3|\Phi_1|^2|\Phi_2|^2\nonumber\\ &+& n_4|\Phi_1^\dagger \Phi_2|^2+[(n_5\Phi_1^\dagger \Phi_2+n_6|\Phi_1|^2\nonumber\\&+& n_7|\Phi_2|^2)(\Phi_1^\dagger \Phi_2)+h.c.] 
\end{eqnarray}
All cofficients are real but for $ m_3,n_5,n_6,n_7 $, which are generally complex.\\\\The VEVs are given by
\begin{eqnarray}\label{019}
<0|\Phi_1 |0>=\left(\begin{array}{c}0\\ \nu_1 \end{array}\right)\: ,\: <0|\Phi_2 | 0>=\left(\begin{array}{c}0\\ \nu_2 e^{i\theta}\end{array} \right)
\end{eqnarray}
where $ \nu_1,\nu_2 $ and $ \theta $ are real. The presence of complex cofficients and the CP-odd phase $ \theta $ make the potential not invariant under CP. \\One of the most important features of these models is flavor changing neutral current processes appearing in Yukawa interactions. As Glashow, Weinberg and Paschos \cite{glashow, paschos} have shown, to suppress these processes the only way to obtain natural flavor conservation is to ensure that only one Higgs doublet has Ykawa interactions with fermions of given charge. To do this, we can impose the following discrete symmetry $ \Phi_1\rightarrow\Phi_1\, , \Phi_2\rightarrow-\Phi_2  $. By this symmetry, we have $ m_3=n_6=n_7=0 $. Therefore CP is conserved in this model although one complex term remains in the potential. We can consider the situation that one allows symmetry $  \Phi_2\rightarrow-\Phi_2 $ to be softly broken by retaining the cofficient $ m_3 $ and $ CP $ is no longer conserved in this state.\\In the next section, we shall describe the EW phase transition, the transition from the symmetric to the broken phase, in the early universe which is strongly first order by adjusing the parameters of the THDM at the critical temperature $ T_c $. After bubbles start to nucleate and expand, they are filled with condensed Higgs fields. In other words, the Higgs VEVs are space-time dependet. The walls of the bubbles are taken a planar which expand along the Z axis. Thus, the Higgs VEVs are given by \cite{nel}  
\begin{align}\label{020}
<0|\Phi_1|0>_T=\frac{\rho_1(z)}{\sqrt{2}}e^{i\theta_1(z)}\:,\: <0|\Phi_2|0>_T=\frac{\rho_2(z)}{\sqrt{2}}e^{i\theta_2(z)} 
\end{align}
 In the symmetric phase, both VEVs vanish, but in the broken phase the VEVs should be close to their zero temperature values. The mass of fermions given by the Yukawa lagrangian is
 \begin{equation}\label{021}
m_\psi(z)=\lambda_\psi\frac{\rho_1(z)}{\sqrt{2}}e^{i\theta_1(z)} 
\end{equation}
where $ \psi $ denotes q or $ \ell $. Therefore, the interaction of a fermion field with the CP-violating Higgs bubble is given by the lagrangian
\begin{align}\label{021}
\cal L_\psi= & \overline{\psi}_Li\gamma^\mu\partial_\mu\psi_L+\overline{\psi}_Ri\gamma^\mu\partial_\mu\psi_R-m_\psi(z)\overline{\psi}_L\psi_R+h.c. 
\end{align}
Equation (\ref{021}) is CP-violating since $ \theta(z) $ cannot be removed by rephasing the fields $ \psi_{L,R} $.
\section{electroweak phase transition}
In this section, we discuss the third sakharov condition, the departure from thermal equilibrium which may is provided by a phase transition. In the unbroken phase the rate of B-violatnig processes, equation (\ref{e012}), is large compared to the expansion rate of the universe which is proportionated to $ T^2 $ in the radiation dominated epoch. Thus these processes can be considered to be in the local thermal equilibrium at these temperatures. Therefore, any baryon asymmetry generated during the phase transition will washed out by the sphaleron-induced processes. If the phase transition is first-order, there are two degenerate thermodynamic states separeted by a barrier at $ T=T_c $ ,and the order parameter, the quantity which undergoes changes when the phase transition occurs, changes discontinuously.\\In the SM, the order parameter is VEV of Higgs field. If the electroweak phase transition is strongly first-order, there are phenomena of bubble nucleation and expansion. The condition that the first-order phase transition is strong in order to aviod being washed out in the broken phase is
\begin{equation}\label{e022}
\frac{\nu _{T_c}}{T_c}\gtrsim 1.
\end{equation}    
The SM cannot provide the EW phase transition to be strongly first-order \cite{buchmuler, kajantie}. Therefore any excess of baryon is washed out by unsuppressed B-violating processes in the broken phase and this cannot explain the baryon asymmetry of the universe. We use the THDM to solve this problem.
\subsection{EW phase transition in the THDM}
The crucial assumption of this paper is that the EWPT occurs in one stage. To do this, we write equation (\ref{e018}) in the Higgs basis. In this basis, only one doublet has VEV, $ \nu $, and the other doublet has zero VEV. By performing the following unitary transformation, we obtain the Higgs basis
\begin{eqnarray}\label{e023}
\left(\begin{array}{c}H_1\\H_2\end{array}\right) &=&\frac{1}{\nu}\left(\begin{array}{cc}\nu_1 &\nu_2 \\ \nu_2 & -\nu_1 \end{array}\right) \left(\begin{array}{c}\Phi_1 \\e^{-i\theta}\Phi_2 \end{array}\right)
\end{eqnarray}
where 
\begin{eqnarray}\label{e024}
H_1&=&\left(\begin{array}{c}0\\(\nu +H)/\sqrt{2} \end{array}\right)\nonumber \\\\ H_2&=&\left(\begin{array}{c}C^+\\(N+iA)/\sqrt{2}\end{array}\right).\nonumber
\end{eqnarray}
Here, H, N, A are the neutral Higgs fields and $ C^{\pm}$ are charged Higgs fields. Thus, the scalar potential in the Higgs basis is
\begin{eqnarray}\label{e025}
V(H_1,H_2)&=&\mu _1|H_1|^2+\mu _2|H_2|^2+(\mu _3 H_1^\dagger H_2+h.c.)\nonumber\\&+&\lambda _1|H_1|^4+\lambda _2|H_2|^4+\lambda _3|H_1|^2|H_2|^2\nonumber\\ &+& \lambda _4|H_1^\dagger H_2|^2+[(\lambda _5H_1^\dagger H_2+\lambda _6|H_1|^2\nonumber\\&+& \lambda _7|H_2|^2)(H_1^\dagger H_2)+h.c.]
\end{eqnarray}
All cofficents are real but for $ \mu_3 ,\lambda_5,\lambda_6,\lambda_7 $, which are generally complex. To suppress flavor changing neutral current, we impose the used discrete symmetry
\begin{equation}\label{e026}
H_1\rightarrow H_1 \: ,\: H_2\rightarrow -H_2.
\end{equation}
This leads to $\mu_3 =\lambda_6=\lambda_7=0  $. For simplicity, we take also $ \lambda_5=0 $, and CP violation is arises from the interaction of fermions with the wall of the Higgs bubble.\\   
We consider H the Higgs boson of the SM and $ N,A,C^{\pm} $ the extra Higgs bosons. By substituting Eq. (\ref{e024}) into Eq. (\ref{e025}) and expanding the potential V in terms of the fields, one obtains the following masses
\begin{align}\label{026}
&m_H^2=\lambda_1 \nu^2 \hspace*{.5cm} ,\hspace*{.5cm} m_{C^\pm}^2=\mu_2 +\frac{1}{2} \nu^2 \lambda_3\nonumber\\ \nonumber \\ &m_N^2=m_A^2=m^2_{C^\pm}+\frac{1}{2} \nu^2\lambda_4 . 
\end{align}
To study the EW phase transition related to the dynamics of the background Higgs field, we consider the effective potential at finite temperatures 
\begin{equation}\label{027}
V_{eff}(H,T)=V(H)+\overline{V}^T_1
\end{equation}
Since the system is in contact with thermal bath in the early universe, we have to consider the finite temperature contributions. The contribution of the one-loop, zero temperature effective potential is expressed as 
\begin{equation}\label{028}
V(H)=V_0+\overline{V}_1
\end{equation}
where $ V_0 $ is the tree level potential and $ \overline{V}_1 $ is the one-loop quantum correction \cite{sher}
\begin{align}\label{029}
V(H)=-\frac{1}{2}(\lambda\nu^2+2D) H^2+\frac{1}{4}\lambda H^4+DH^4\ln(H^2/\nu^2) 
\end{align}
here, $ m_H^2=\lambda_1\nu^2=(2\lambda\nu^2+12D) $ , $\lambda $ is one-loop quartic scalar coupling and
\begin{align}\label{e030}
D=\frac{1}{64\pi^2 \nu^2}(6m_W^4+3m_Z^4-12m_t^4+2m_{C^\pm}^4+m_N^4+m_A^4).
\end{align}
Formalism applied at finite temperature is analogous to the formalism used at zero temperature, but the crucial point is that the familiar boundary conditions $ t=\pm\infty $
are relevant at zero temperature and periodic boundary conditions are relevant at finite temperature. Therefore, path integral may also used at finite temperature with appropriate replacement of the free 2-point function of zero temperature with one of finite temperature. The finite temperature contribution is given by
\begin{eqnarray}\label{e031}
\overline{V}^T_1=\frac{T^4}{2\pi^2}\sum_B n_B \int_0^\infty dx\, x^2 \ln(1-e^{-\sqrt{x^2+\beta^2m_B^2}})\nonumber\\-\frac{T^4}{2\pi^2}\sum_F n_F \int_0^\infty dx\, x^2 \ln(1+e^{-\sqrt{x^2+\beta^2m_F^2}}) 
\end{eqnarray}
where $ n_{B(F)} $ denotes the number of degrees of freedom: $ n_W=6, n_Z=3, n_t=-12,n_{C^\pm}=2, n_N=n_A=1 $ moreover $ \beta=1/T $ and $ m_{B(F)} $ denotes the mass of a boson(fermion) in the presence of the bachground field H.\\In the high temperature limit, when $ m(H)/T $ is small, $ \overline{V}_1^T $ may be expanded \cite{dolan}:
\begin{align}\label{e032}
\overline{V}_1^T&=\sum_B n_B\left[\frac{m_B^2 T^2}{24}-\frac{m_B^3 T}{12\pi}-\frac{m_B^4}{64\pi^2}\ln\frac{m_B^2}{T^2}+\frac{c_B m_B^4}{64\pi^2}\right]\nonumber\\&+\sum_F n_F\left[\frac{m_F^2 T^2}{48}+\frac{m_F^4}{64\pi^2}\ln\frac{m_F^2}{T^2}+\frac{c_F m_F^4}{64\pi^2}\right]+O(\frac{m^6}{T^2})
\end{align}
 where $ c_B=2\ln 4\pi-2\gamma  $, $ c_F=2\ln \pi-2\gamma  $ and $ \gamma $ is Euler constant. The field-dependent mass of the particle of the theory may written as $ m^2(H)\simeq m^2 H^2/\nu^2 $. The high-temperature approximation agrees with exact potential for $ m/T<1.6(2.2) $ for fermons(bosons) \cite{anderson}.\\Finally, the finite temperature effective potential is given by
\begin{align}\label{e033}
&V_{eff}(H,T)\simeq  A(T^2-T^2_*)H^2-BTH^3+\frac{\lambda_T H^4}{4}+\cdots
\end{align}\label{e033}
where,
\begin{align}\label{ex033}
 &A=\frac{1}{24\nu^2}(6m_W^2+3m_Z^2+6m_t^2+2m_{C\pm}^2+m_N^2+m_A^2),\\ &T_*^2=\frac{m_H^2}{4A}-\frac{2D}{A}=\frac{1}{A}(\frac{m_H^2}{4}-2D),\\& B=\frac{1}{12\pi\nu^3}(6m_W^3+3m_Z^3+2m_{C^\pm}^3+m_N^3+m_A^3),\\&\lambda_T=\lambda_1-\frac{1}{16\pi^2\nu^4}[6m_W^4(\ln\frac{m_W^2}{T^2}-c_B)\nonumber\\&+3m_Z^4(\ln\frac{m_Z^2}{T^2}-c_B)-12m_t^4(\ln\frac{m_t^2}{T^2}-c_F)\nonumber\\&+4m_i^4(\ln\frac{m_i^2}{T^2}-c_B)].  
\end{align}
The cubic term causes the phase transition to be first-order.% As we said before, the condition that the phase transition is strongly first-order in order to suppress the B-violating processes in the broken phase is $ \nu_{T_c}/T_c\gtrsim 1 $ % 
\begin{figure}[th]
\begin{center}
\includegraphics[width=8cm, height=7cm]{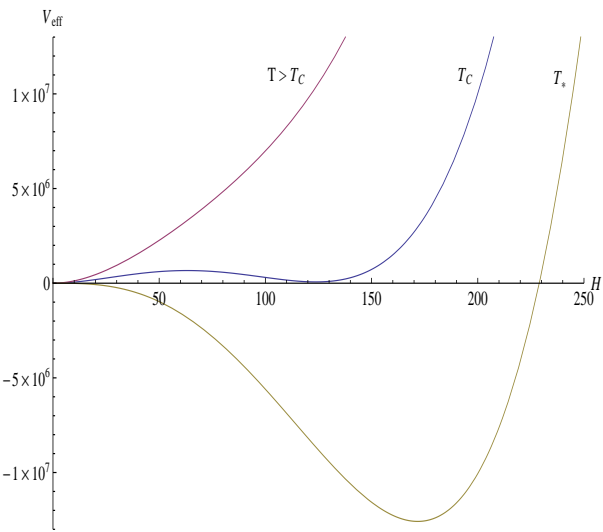}\caption{\label{f1} The first-order phase transition for the effctive potential and the various behavior of the effective potential before and after the phase transition, at temperatures $ T=133\,GeV > T_c $ and $ T_*=115\,GeV $, is shown. The condition of the strong phase transition $ \nu_{T_c}=T_c $ is imposed. }
\end{center}
\end{figure} 
 As seen from Fig. \ref{f1}, the EW phase transition temperature, $ T_c $, is the temperature at which there are two degenerate states. In other words, the free energies of the symmetric and broken phases are equal. Thus, the second minimum which is degenerate with $ \nu_T=0 $ is the other root of $ V_{eff} $. By dividing $ V_{eff} $ by $ H^2$, two same roots are $ \nu_{T_c}=2BT_c/\lambda_T$ where
 \begin{eqnarray}\label{e034}
 T_c=\frac{T_*}{\sqrt{1-\frac{B^2}{\lambda_ T A}}}. 
 \end{eqnarray}    
 At temperatures above $ T_c $, $ V_{eff} $ has one minimum in $ \nu_T=0 $. As the temperature falls, at temperature $ T'=T_*/\sqrt{1-9B^2/8\lambda_ T A} $ , the potential obtains a local minimum $ H_{T'}=3BT'/2\lambda_{T'} $. At lower temperatures, this point is separated into a local minimum $ H_{T_-} $ and a local maximum or a barrier $ H_{T_+} $ 
 \begin{eqnarray}\label{e035}
 H_{T_\pm}=\frac{3BT}{2\lambda_T}\left[1\pm \sqrt{1-\frac{8A\lambda_T}{9B^2}(1-\frac{T_*^2}{T^2})}\right]. 
 \end{eqnarray} 
 After the phase transition, bubbles filled with the Higgs condensate start to nucleate. At the temperature $ T_* $, $ H_{T_-}=0 $ and $ \nu_{T_*}=3BT_*/\lambda_{T_*} $ . Therefore, at this temperature bubbles fill all of the volume.\\  As we said before, the condition that the phase transition is strongly first-order, in order to suppress the B-violating processes in the broken phase, is $ \nu_{T_c}/T_c\gtrsim 1 \Rightarrow 2B=\lambda_{T_c} $. If we assume $ \lambda_4 \rightarrow 0 $, we obtain $ m_i=m_{c^\pm}=m_N=m_A  $.
 \begin{figure}[th]
 \begin{center}
 \includegraphics[width=7.5cm, height=7.5cm]{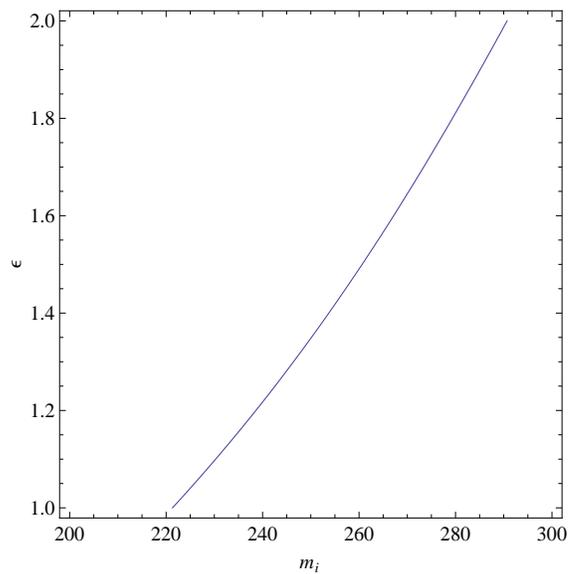}\caption{ \label{f2} Defined by $ \epsilon=\nu_{T_c}/{T_c} $, The vertical axis is the phase transition strength and the horizontal axis is the mass of the extra Higgs bosons.} 
 \end{center}
 \end{figure}
 
 Taking $ \nu=246$ GeV, $m_W=80$ GeV, $m_Z=91 GeV$, and $ m_t=173$ GeV \cite{beringer} , condition (\ref{e022}) may be satisfied for $ m_H=126$ GeV  provided the mass of the extra Higgs bosons is about  221 GeV. Then, by using equation (\ref{e022}) we obtain the critical temperature $ T_c\simeq 124\,GeV  $. The strength of the phase transition is defined by $ \epsilon =\nu_{T_c}/T_c $. If we insert $ T_c $ into $ \lambda_T $, we can obtain a relation between $ \epsilon $ and $ m_i $ in which $ \epsilon $ increases up to very strong phase transition as the mass of the extra Higgs bosons grows ( see Fig. \ref{f2} ).
\section{electroweak barogenesis} 
\begin{figure}[th]
\begin{center}
\includegraphics[width=6.5cm, height=6.5cm]{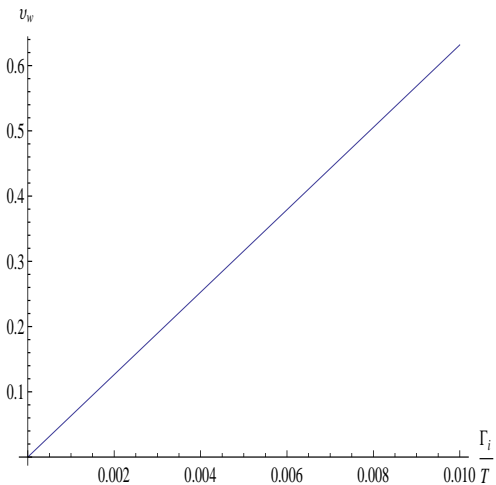}\caption{\label{f3} The wall velocity as a function of $ \Gamma_i /T $ is depicted. The greater the friction coefficient is, the less the wall velocity will be.}
\end{center}
\end{figure}

 In the previous sections, we expressed the conditions that are necessary to produce baryons. Now, we discuss how these conditions can be used to describe observed asymmetry. There are two mechanisms for EW baryogenesis when bubble walls pass through space:
 \begin{enumerate}
 \item Local baryogenesis: in which B-violating processes and CP-violation together occur on or at the bubble walls.
 \item Nonlocal baryogenesis: in this case, particles and antiparticles interact with a CP-violating bubble wall. This cause an asymmetry in a quantum number other than baryon number and the asymmetry is transformed into the asymmetry in the baryon number by sphaleron processes. There are two calculation regimes in the nonlocal barogenesis. If scattering effects is not considered or in other words the mean free path of the fermions, $ l_f $, is larger than the thickness of the wall, $ l_w $, this regime is referred to the thin wall regime.\\On the other hand, if the mean free path of the fermions is less than or the same order as the wall thichness, then the scattering effects should be considered which it is called the thick wall regime.\\ 
 \end{enumerate} 
 The essential quantities of the bubble wall dynamics are the wall velocity $ \upsilon_w  $ and thickness $ l_w $. To anatically calculate these quantites, we use thin wall limit (for detail see \cite{wall, mege1, mege2}). In this approximation free energy of the bubble is given by
 \begin{eqnarray}\label{e36}
 \mathcal{F}=\frac{4\pi}{3}R^3\Delta p(T)+4\pi R^2\sigma(T) 
 \end{eqnarray}
 where $ \Delta p\equiv V_{eff}(\nu_T,T)-V_{eff}(0,T)  $ is the pressure difference between the two sides of the wall, which it is necessary to grow the bubbles. $ \sigma (T) $ is surface tension of the wall,
 \begin{eqnarray}\label{e37}
 \sigma=\int (\frac{d H}{d r})^2 dr =\int_0^{\nu_T}\sqrt{2V_{eff}}d H 
 \end{eqnarray}
 $ \sigma $ change insignificantly within phase transition, therefore surface tension can be estimated by,
 \begin{eqnarray}\label{e38}
 \sigma(T_c)=\frac{2\sqrt{2}B^3T_c^3}{3\lambda_{T_c}^{5/2}}. 
 \end{eqnarray}
  Maximizing with respect to $ R $, we obtain free energy of the critical bubble
 \begin{eqnarray}\label{e37}
 \mathcal{F}_c=\frac{16\pi\sigma^3}{3\Delta p^2}. 
 \end{eqnarray} 
 By inverting the equation $ \frac{d H}{dr}=-\sqrt{2V_{eff}}  $ one can obtain the wall thickness. We roughly obtain
 \begin{eqnarray}\label{e39}
 l_w\sim \frac{\nu_{T_c}}{\sigma(T_c)}\sim 0.01. 
 \end{eqnarray}
 In the baryogenesis context, by ignoring hydrodynamics the wall velocity is given by \cite{mege1}
 \begin{eqnarray}\label{e40}
 \upsilon_w\sim \frac{\Delta p}{\eta},  
 \end{eqnarray}
 where $ \eta $ is the friction coefficient. In thin wall approximation we apply the linear approximation for $ \Delta p\equiv \Delta V_{eff}\simeq \frac{L(T_c-T)}{T_c}  $, therefore we have
 \begin{eqnarray}\label{e41}
 \upsilon_w\sim \frac{L(T_c-T)}{\eta T_c}, 
 \end{eqnarray}
 where latent heat is given by,
 \begin{align}\label{e42}
 L&=T_c(\frac{d V_{eff}(\nu_T,T_c)}{d T}-\frac{d V_{eff}(0,T_c)}{d T})\nonumber\\&=8A(T_c T_*)^2(\frac{B}{\lambda_{T_c}})^2. 
 \end{align}
 The friction coefficient is given by $ \eta=\eta_t+\eta_i $ where $ \eta_t $ is the contribution of particles with a thermal distribution and $ \eta_i $ is the contribution of infrared bosons and are given by \cite{mege2}
 \begin{eqnarray}\label{e43}
 \eta_t &=&\sum_i \frac{g_i\lambda_i^4T}{\Gamma_i}(\frac{\ln \chi_i}{2\pi^2})^2\frac{H^2\sigma}{T},\\
\eta_i &=&\sum_b\frac{g_i m_D^2T}{32\pi l_w}\ln(m_i(H)l_w),
 \end{eqnarray}
 where $ g_i $ is the number of degrees of freedom of species i, $ m_i(H)=\frac{Hm_i}{\nu} $, for fermions $ \chi_i=2 $ and for bosons $ \chi_i=m_i(H)/T $, $ m_D\sim \lambda_i T $ is the Debye mass and interaction rates are $ \Gamma_i/T $. Finally, we can calculate the velocity at the nucleation temperature $ T_N \simeq T_c $,
 \begin{eqnarray}\label{e44}
 \upsilon_w(T_N)&\sim & \frac{1}{\eta}(\frac{16\pi \sigma^3}{3T_c})^{1/2}(4\ln\frac{M_0}{T_c})^{-1/2}, 
 \end{eqnarray}
 where $ M_0\sim 10^{18}\,GeV $. This equation is obtained from calculations of the relevant time scale for the formation of a critical bubble. As seen from Fig. \ref{f3}, for the given interaction rates \cite{mege2, joyce96a} we find the velocities of the wall which are less than the sound speed in relativistic plasma. Therefore, we can apply nonlocal baryogenesis in this case.
Also, the mean free path of fermions is roughly given by: $ l_f \sim (16 \pi\alpha^2T)^{-1} $ where $ \alpha $ is the coupling strength. Hence, considering equation (\ref{e39}) and strong sphaleron effects which act on the quarks \cite{giu} the main sources of baryogenesis are the leptons in the thin wall regime. Therefore, they can be taken as free particles interacting with Higgs field in the small region. Since the Yukawa coupling of the tau lepton is larger than other leptons, the main contribution of the interactions with wall belongs to this particle.\\The lagrangian of the fermions in the background of a CP-violating Higgs bubble wall is \cite{joyce96a, joy} 
 \begin{align}\label{e036}
 L_\psi= & \overline{\psi}i\gamma^\mu[\partial_\mu +i(\frac{\frac{1}{2}\upsilon_2^2\partial_\mu (\theta _1-\theta _2)+\upsilon_1^2\partial_\mu\theta_1 +\upsilon_2^2\partial_\mu\theta_2}{\upsilon_1^2+\upsilon_2^2})\gamma^5]\nonumber\\ & \psi -m\overline{\psi}\psi . 
\end{align}
Now, in the given THDM where $ \nu_2=0, \nu_1=\nu, \theta_2=0, \theta_1=\theta $ this equation reduces to
 \begin{eqnarray}\label{e037}
 L_\psi=\overline{\psi}i\gamma^\mu[\partial_\mu +i(\partial_\mu \theta)\gamma^5]\psi-m\overline{\psi}\psi . 
 \end{eqnarray}
 To calculate the reflection probabilities of particle and antiparticles with a given chirality, we use the equation of motion in the quantum mechanics case. From the lagrangian (\ref{e037}) in the rest frame of the wall, we obtain \cite{nel}
 \begin{eqnarray}\label{e038}
 i\partial_z \left(\begin{array}{c} \psi_1\\ \psi_3 \end{array}\right)=\left(\begin{array}{cc} E' & m^*\\-m & -E'\end{array}\right) \left(\begin{array}{c} \psi_1 \\ \psi_3\end{array}\right),
 \end{eqnarray} 
 \begin{eqnarray}\label{e039}
  i\partial_z \left(\begin{array}{c} \psi_2\\ \psi_4 \end{array}\right)=\left(\begin{array}{cc} E' & m\\-m^* & -E'\end{array}\right) \left(\begin{array}{c} \psi_2 \\ \psi_4\end{array}\right),
 \end{eqnarray}
where $ \psi_1, \psi_2 $ correspond to an incident right-handed and left-handed particle, respectively, in the symmetric phase. $ \psi_3, \psi_4 $ correspond to a left-handed and right-handed particle, respectively, reflected into the symmetric phase. $ E' $ is the energy of the incident particle in the rest frame of the wall and $ m=m(z) e^{i\Delta\theta} $.\\ Conservation of electric charge forbids the change of charged particles into antiparticles during the reflection process. furthermore, conservation of angular momentum mandates that a left-handed fermion is reflected into right-handed one and vice versa since the linear momentum changes. Due to energy conservation, all of fermions incident from the symmetric phase with the energy less than $ m_f $ are completely reflected and the fermions with the energy more than $ m_f $ have non-zero coefficients of the reflection  and transition. \\The probability that a right-handed fermion is reflected into a left-handed fermion is equal to the probability that a left-handed antifermion is reflected into a right-handed antifermion unless CP is violated which it can be seen via the analytic calculation of Joyce et al. \cite{joyce94b}, 
\begin{align}\label{e040}
\Delta R\equiv & R_{R\rightarrow L}-R_{\overline{L}\rightarrow \overline{R}}\nonumber\\=&\frac{4t(1-t^2)}{|m_f|}\int_{-\infty} ^\infty dz \,\rm{Im} [m \, m_f^*]\cos(2p_zz), 
\end{align}
where $ t=\tanh\omega $ and $ 2t\thickapprox \tanh2\omega=\frac{m_f}{p_z} $, $ m_f\equiv m(-\infty) $ is the mass of the fermion in the broken phase and $ p_z $ is the momentum of the fermion at infinity, in the broken phase. This equation is valid for the incident momenta range $ m_f<|p_z|<\frac{1}{l_w} $ . Finally,
\begin{eqnarray}\label{e041}
\Delta R\simeq\frac{2m_f^2}{p_z m_H}\sin\Delta\theta .
\end{eqnarray}
CPT invariance implies:
\begin{align}\label{e042}
&R_{L\rightarrow R}=R_{\overline{L}\rightarrow\overline{R}} \hspace*{.5cm} ,\hspace*{.5cm} R_{R\rightarrow L}=R_{\overline{R}\rightarrow\overline{L}}\\\nonumber\\
&\Rightarrow R_{R\rightarrow L}-R_{\overline{L}\rightarrow\overline{R}}=R_{\overline{R}\rightarrow\overline{L}}-R_{L\rightarrow R}. 
\end{align} 
Therefore, the difference between the fluxes of the left-handed fermions and right-handed antifermions, $ J^L_f $, injected from the wall into the symmetric phase equals to $ J_f^R $, which is the difference between the fluxes of the right-handed fermions and left-handed antifermions. This results in the net baryon number to be zero.\\Nevertheless, the weak sphaleron processes act only on the left-handed fermions and right-handed antifermions in the symmetric phase. Thus, $ J_f^L>0 $ makes left-handed fermions to be produce more than right-handed antifermions. Then, the excess baryon number is swept  by the bubble and this value persist since the weak sphaleron processes is suppressed in the broken phase.\\To express the net flux injected into symmetric phase, we also should account for the transmission probability of the particles injected from broken phase. According to CPT invariance, we can find the following relations between the transmission and reflection probabilities
\begin{eqnarray}\label{e043}
T_{R\rightarrow R}=1-R_{R\rightarrow L} =T_{\overline{R}\rightarrow\overline{R}}=1-R_{\overline{R}\rightarrow \overline{L}},\\\nonumber\\ 
T_{L\rightarrow L}=1-R_{L\rightarrow R}= T_{\overline{L}\rightarrow\overline{L}}=1-R_{\overline{L}\rightarrow \overline{R}}.
\end{eqnarray}   
The contribution of reflected particles to the symmetric phase include $ \Delta R f_s $ and that of transmited particles include $ (T_{\overline{R}\rightarrow\overline{R}}-T_{L\rightarrow L})f_b=-\Delta R f_b $ where $ f_s $ and $ f_b $ are the phase-space densities of the particles and antiparticles in the symmetric and broken phase in the wall frame, respectively.\\The flux of the injected left-handed fermions is given by
\begin{eqnarray}\label{e044}
J_f^L=\int_{p_{z<0}} \frac{d^3p}{(2\pi)^3}\frac{|p_z|}{E}(f_s-f_b)\Delta R. 
\end{eqnarray}
If the wall were at rest, $ (f_s-f_b) $ as well as the net baryon number would vanish. Using equation (\ref{e041}) and taking $ (f_s-f_b) $ to linear order in $ \upsilon_w $ and leading order in $ m_f/m_H ,m_H/T $, we obtain
\begin{eqnarray}\label{e045}
J_f^L=-J_f^R=\frac{\upsilon_w}{4\pi^2} m_f^2 m_H \sin\Delta\theta . 
\end{eqnarray} 
The rate of production of baryons is given by \cite{joyce96a}
\begin{eqnarray}\label{e046}
\frac{dn_B}{dt}=-\frac{n_f \Gamma(T)}{2T}\sum_i \mu_i ,
\end{eqnarray}
where $ \Gamma(T) $ is given by equation (\ref{e012}). $ n_f $ is the number of families and $ \mu_i $ is the chemical potential of the left-handed particles of species i. If there is local thermal equilibrium in front of the bubble wall, then the relation between the chemical potential $ \mu_i $ of particles and their number density $ n_i $ is
\begin{eqnarray}\label{e047}
n_i=\frac{T^2}{12}\mu_i .
\end{eqnarray}
Now, using the continuity equation, we write the diffusion equation in the symmetric phase \cite{joyce96a} 
\begin{eqnarray}\label{e048}
\frac{d n_i}{dt}=\frac{\partial n_i}{\partial t}+\vec{\nabla}\cdot\vec{J_i}=-\sum_i\frac{\Gamma(T)}{2T}\mu_i , 
\end{eqnarray}
where $ \vec{J_i}=-D_i\vec{\nabla}n_i+\vec{J}^{inj}_i $, $ D_i $ is the diffusion constant and $ \vec{J_i^{inj}}=\xi_i J_i \delta(z-\upsilon_w t) $  where $ J_i $ is the net reflected flux of species i, for example \rm Eq. (\ref{e045}). $ \xi_i $ is the persistance length of the current in front of the wall where it contains all uncertainties on how the injected flux thermalizes in the symmetric phase. Assuming that the persistance length of current is much larger than the wall thickness, we can approximate the injected current with a delta function. One can solve the equation (\ref{e048}) for leptons. Also, we look for stationary solutions in the the wall frame, $ n(z,t)\equiv n(z-\upsilon_w t) $. The decay time of leptons is longer than the time which the wall spends to pass and we can neglet decays. Thus 
\begin{eqnarray}\label{e049}
D_L l''_L+\upsilon_w l'_L=\xi_L J^L_l\delta' 
\end{eqnarray}
where prime denotes the spatial derivative in the Z direction. Requiring that the solutions be zero at $ \infty $ we shall obtain
\begin{eqnarray}\label{e050}
l_L(z)=&J_l^L \frac{\xi_L}{D_L}e^{\frac{-\upsilon_w }{D_L}z}\hspace*{1cm} z>0.
\end{eqnarray}
Using equations (\ref{e046}) and (\ref{e047}) in the wall frame, we obtain the baryon number density on the wall
\begin{eqnarray}\label{e051}
n_B=-\frac{6n_f D_L}{\upsilon_w^2 T^3}J^L_l\frac{\Gamma(T)\xi_L}{D_L}. 
\end{eqnarray}
We must divide $ n_B $ by the entropy density, $ \frac{2\pi^2}{45}g_* T^3 $ where $ g_*\thickapprox 100 $ the number of relativistic degrees of freedom, 
\begin{align}\label{e052}
\frac{n_B}{s}=\frac{\xi_L}{D_L}\frac{45}{4\pi^4 g_* \upsilon_w}(\frac{m_\tau}{T})^2 (\frac{m_H}{T}) \frac{6n_f\Gamma(T)}{T^3}D_L\sin\Delta\theta , 
\end{align}
where $ m_f $ is the tree-level finite temperature mass, $ m_f\sim \lambda_f T $. Finally, puting the numerical values in the above equation, $ \frac{\xi_L}{D_L}\sim \frac{m_H}{T} $, $ D_L\sim 1 $ \cite{joyce96a}, $ \upsilon_w\sim 0.1 $ and $ \lambda_\tau=0.01 $ in $ T=124 $, we obtain $ \frac{n_B}{s}\sim 7\times 10^{-10} $ if $ \sin\Delta\theta \sim 1 $.\\\\

\section{conclusion}
In this paper we assume that the production of baryons occur at the temperatures of order $ 100 $ GeV. We presented an analytic calculation of EW phase transition within the context of the two-Higgs-doublet model. Our proposed phase transition is strongly first-order and occurs in one stage. This leads to the prediction that the mass of the extra Higgs-bosons about $ 221\,GeV $ and SM Higgs-boson mass is $ 126\,GeV $. We calculate the wall velocity as $ 0.1\lesssim \upsilon_w \lesssim  0.6  $ and  the wall thickness as $ l_w\sim 0.01 $. Finally by nucleation of thin-walled bubbles and using nonlocal baryogenesis, we obtain $ \frac{n_B}{s}\sim 7\times 10^{-10} $ which is comparable with observed value.

\begin{acknowledgments} \label{Calculation}
I would like to thank Siamak. S. Gousheh for useful discussions.
\end{acknowledgments}

 \end{document}